\documentclass[aip,jcp,reprint,noshowkeys,superscriptaddress]{revtex4-2}

\usepackage{graphicx,dcolumn,bm,xcolor,microtype,multirow,amscd,amsmath,amssymb,amsfonts,physics,wrapfig,bbold,siunitx,xspace,tikz}
\usepackage[version=4]{mhchem}
\usepackage[utf8]{inputenc}
\usepackage[T1]{fontenc}
\usepackage[caption=false]{subfig}
\usepackage{hyperref}
\usepackage{tabularray}
\DefTblrTemplate{caption-tag}{default}{\tablename~\thetable.}
\DefTblrTemplate{caption-sep}{default}{\space}
\DefTblrTemplate{caption}{default}{%
  \UseTblrTemplate{caption-tag}{default}%
  \UseTblrTemplate{caption-sep}{default}%
  \UseTblrTemplate{caption-text}{default}%
}
\DefTblrTemplate{capcont}{default}{%
  \UseTblrTemplate{caption-tag}{default}%
  \UseTblrTemplate{caption-sep}{default}%
  \UseTblrTemplate{caption-text}{default}
  \UseTblrTemplate{conthead-text}{default}%
}
\SetTblrStyle{conthead}{
  halign = j,
}
\hypersetup{
	colorlinks,
	linkcolor={purple},
	citecolor={purple},
	urlcolor={purple}
}
\usepackage{listings}
\definecolor{codegreen}{rgb}{0.58,0.4,0.2}
\definecolor{codegray}{rgb}{0.5,0.5,0.5}
\definecolor{codepurple}{rgb}{0.25,0.35,0.55}
\definecolor{codeblue}{rgb}{0.30,0.60,0.8}
\definecolor{backcolour}{rgb}{0.98,0.98,0.98}
\definecolor{mygray}{rgb}{0.5,0.5,0.5}

\definecolor{sqred}{rgb}{0.85,0.1,0.1}
\definecolor{sqgreen}{rgb}{0.25,0.65,0.15}
\definecolor{sqorange}{rgb}{0.90,0.50,0.15}
\definecolor{sqblue}{rgb}{0.10,0.3,0.60}

\lstdefinestyle{mystyle}{
	backgroundcolor=\color{backcolour},
	commentstyle=\color{codegreen},
	keywordstyle=\color{codeblue},
	numberstyle=\tiny\color{codegray},
	stringstyle=\color{codepurple},
	basicstyle=\ttfamily\footnotesize,
	breakatwhitespace=false,
	breaklines=true,
	captionpos=b,
	keepspaces=true,
	numbers=left,
	numbersep=5pt,
	numberstyle=\ttfamily\tiny\color{mygray},
	showspaces=false,
	showstringspaces=false,
	showtabs=false,
	tabsize=2
}

\newcolumntype{d}{D{.}{.}{-1}}

\DeclareSIUnit{\au}{a.u.}
\DeclareSIUnit{\debye}{D}

\usepackage[normalem]{ulem}

\newcommand{\SupMat}{\textcolor{blue}{Supporting Information}\xspace}

\newcommand{\fnm}{\footnotemark}
\newcommand{\fnt}{\footnotetext}

\newcommand{\br}{\boldsymbol{r}}

\newcommand{\Ns}{{N_s}}

\DeclareMathOperator{\diag}{diag}

\newcommand{\bO}{\boldsymbol{0}}
\newcommand{\bI}{\boldsymbol{1}}
\newcommand{\bA}{\boldsymbol{A}}
\newcommand{\bB}{\boldsymbol{B}}

\newcommand{\bT}{\boldsymbol{T}}

\newcommand{\bH}{\boldsymbol{H}}

\newcommand{\bX}{\boldsymbol{X}}
\newcommand{\bY}{\boldsymbol{Y}}
\newcommand{\bZ}{\boldsymbol{Z}}

\newcommand{\bOmega}{\boldsymbol{\Omega}}

\newcommand{\abraket}[2]{\braket*{#1\vert}{#2}}

\newcommand{\ii}{\text{i}}

\newcommand{\upar}{\uparrow}
\newcommand{\dwar}{\downarrow}
\newcommand{\LCPQ}{Univ Toulouse, CNRS, Laboratoire de Chimie et Physique Quantiques, Toulouse, France}
\begin{document}	
\title{RPAx Instabilities from Ground- and Excited-State References}
\author{Loris \surname{Burth}}
\email{lburth@irsamc.ups-tlse.fr}
\affiliation{\LCPQ}
\author{Marios-Petros \surname{Kitsaras}}
\email{kitsaras@irsamc.ups-tlse.fr}
\affiliation{\LCPQ}
\author{Pierre-Fran\c{c}ois \surname{Loos}}
\email{loos@irsamc.ups-tlse.fr}
\affiliation{\LCPQ}
\begin{abstract}
Random-phase approximation with exchange (RPAx) calculations based on ground- or excited-state references can yield complex excitation energies, causing the corresponding RPAx correlation energy to become complex as well. 
In this work, we investigate the origin of these complex energies using \ce{H2} in the 6-31G basis set as a minimal model. 
For the ground-state restricted Hartree-Fock reference, complex roots emerge at the well-known triplet instability associated with the appearance of a lower-lying unrestricted, symmetry-broken solution. 
For excited-state references, we identify two distinct mechanisms leading to complex excitation energies. 
The first is similarly associated with symmetry breaking. 
The second arises when a positive-energy excitation becomes degenerate with the positive-energy de-excitation associated with a negative-energy transition.
Finally, we demonstrate the practical implications of these findings for spin-flip Bethe-Salpeter equation calculations of cyclobutadiene and state-specific core-ionization calculations for selected systems. 
These examples show that, in the absence of the relevant instabilities, methods sharing the generalized eigenvalue structure of RPAx can be applied to excited-state references without resorting to the Tamm-Dancoff or core-valence separation approximations, while retaining real excitation energies.
  \bigskip
\begin{center}
	\boxed{\includegraphics[width=0.5\linewidth]{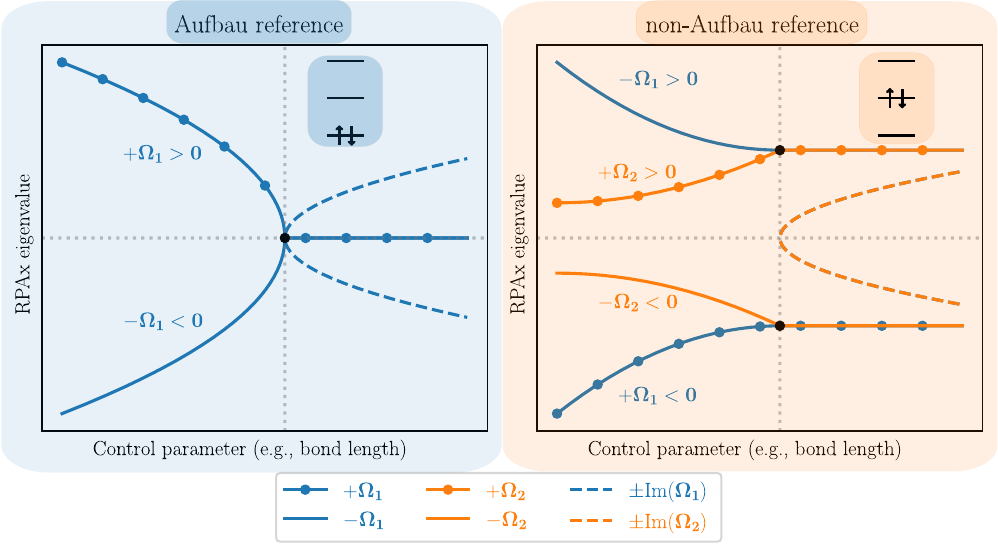}}
\end{center}
\bigskip
\end{abstract}
\maketitle
\section{Introduction}
\label{sec:introduction}
\begin{figure*}
	\centering
	\includegraphics[width=0.6\textwidth]{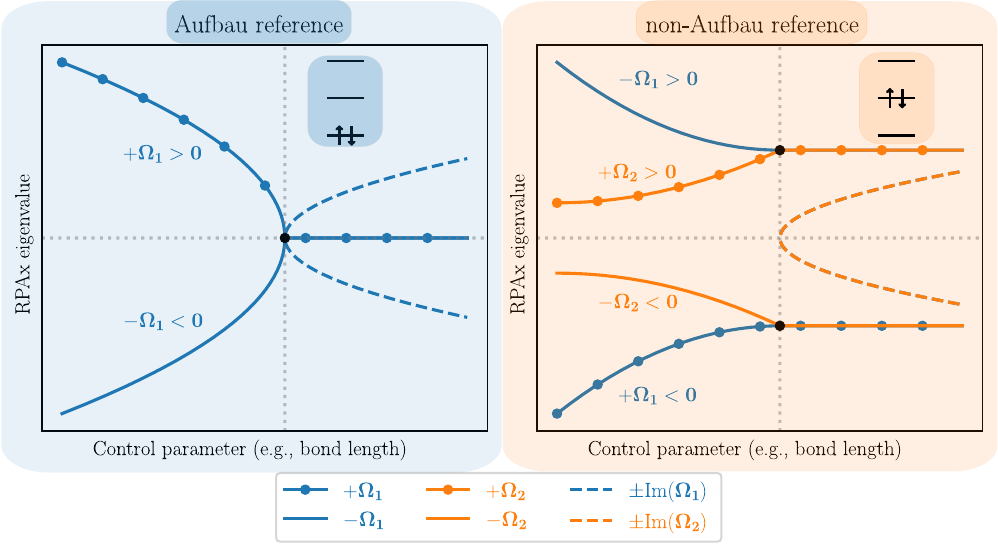}
  \caption{
	Schematic representation of the two mechanisms leading to complex RPAx excitation energies.
	Left: for an Aufbau reference, an excitation and its de-excitation counterpart, $+\Omega_1$ and $-\Omega_1$, coalesce at zero energy, signaling a conventional instability of the reference and subsequently giving rise to a pair of purely imaginary eigenvalues.
	Right: for a non-Aufbau reference, a positive-energy excitation $+\Omega_2$ can become degenerate with the positive-energy de-excitation $-\Omega_1$ associated with a negative-energy excitation $+\Omega_1$.
	The corresponding negative-energy partners become degenerate simultaneously, and beyond the degeneracy the two pairs acquire nonzero imaginary components.
	Solid lines indicate real eigenvalues, while dashed lines indicate the corresponding imaginary branches.} 
	\label{fig:scheme}
\end{figure*}
The accurate prediction of electronic excitation and ionization energies is essential for understanding photochemical processes, material properties, and spectroscopic measurements.
Computationally affordable descriptions commonly begin with self-consistent-field (SCF) methods such as Hartree-Fock (HF)~\cite{Hartree_1928,Fock_1930,Roothaan_1951} and density-functional theory (DFT).~\cite{Kohn_1965,Seidl_1996}
Within these approaches, ionization energies can be estimated using the conventional or generalized Koopmans' theorem.~\cite{Koopmans_1934,Szabo_1996}

Excited and ionized states can be treated more explicitly in several ways.
One possibility is to perform state-specific SCF calculations by selecting the occupied orbitals using the maximum overlap method (MOM).~\cite{Gilbert_2008,Barca_2018,Corzo_2021}
The excitation energies are then evaluated from differences between total energies.
Additionally, dynamical correlation can be introduced through wavefunction methods such as configuration interaction (CI) or coupled-cluster (CC) theory.
Linear-response approaches provide another route to excitations and ionizations.
These include time-dependent DFT, time-dependent HF (TDHF),~\cite{Gell_Mann_1957,McLachlan_1964} and equation-of-motion CC (EOM-CC) theory.~\cite{Emrich_1981,Sekino_1984,Geertsen_1989}
Hereafter, we refer to TDHF as the random-phase approximation with exchange (RPAx).~\cite{McLachlan_1964,Hesselmann_2011}

RPA methods are closely connected to CC with double excitations (CCD).
Particle-particle RPA is equivalent to ladder CCD,~\cite{Scuseria_2013,Peng_2013} whereas particle-hole RPAx is related to ring CCD.~\cite{Scuseria_2008,Rekkedal_2013}
These formalisms provide access to doubly charged and neutral excitations, respectively, from a single-determinant reference.
Unlike Hermitian CI problems, however, the RPA eigenvalue problem has a non-Hermitian structure and may possess complex eigenvalues.
For a ground-state HF reference, the real-valued nature of the RPAx excitation energies is intimately connected to the stability of the underlying determinant.
A loss of positive semidefiniteness of the HF stability matrix may produce complex excitation energies and, consequently, a complex RPAx correlation energy.~\cite{Scuseria_2013,Song_2025}
In the left panel of Fig.~\ref{fig:scheme}, we illustrate that this is related to a vanishing RPAx excitation energy before it becomes purely imaginary.

Complex energies are not unique to RPA.
They have also been observed in the non-Hermitian EOM-CC methods, for example, near conical intersections and degeneracies between states of the same symmetry.~\cite{Koehn_2007,Kjonstad_2017,Kjonstad_2024,Plasser_2014,Fleig_2007,Thomas_2021, Grazioli_2025}
In that context, similarity-constrained CC approaches have been developed to remove such artifacts by introducing additional flexibility into the wavefunction and excitation operator, thus enforcing the linear independence of the intersecting states.~\cite{Kjonstad_2017a,Kjonstad_2019,Stoll_2025}
Another way to avoid complex energies in the context of CC methods is using unitary CC.~\cite{Grazioli_2025}

The situation is more subtle when the reference determinant represents an excited state.
Since an excited-state HF solution is in general not a minimum but a stationary point, its response spectrum contains transitions toward both higher- and lower-lying states.
In EOM-CC theory, this feature is exploited in spin-flip calculations, where a high-spin triplet reference provides access to higher- and lower-lying singlet states, including states with substantial multiconfigurational character.~\cite{Casanova_2020,Krylov_2001a,Krylov_2001,Levchenko_2004,Manohar_2008} 
Analogous spin-flip formulations have been developed for RPA and the Bethe-Salpeter equation (BSE) formalism.~\cite{Sears_2003,Casanova_2012,Li_2012,Casanova_2020,Monino_2021}

For an excited-state reference, the RPAx response problem can give rise to complex excitation energies due to the coupling between excitation and de-excitation modes.
This is illustrated in the right panel of Fig.~\ref{fig:scheme}.
A positive-energy excitation can become degenerate with the positive-energy de-excitation associated with a negative-energy transition.
When these modes coalesce, their eigenvalues can become complex.
Two decoupling approximations are commonly employed to address these difficulties.
The Tamm-Dancoff approximation (TDA) neglects the coupling between excitations and de-excitations, converting the RPAx problem into the Hermitian CI with singles (CIS) problem.~\cite{Fetter_1971,Ring_1980}
The core-valence separation (CVS) approximation, applied to a non-Aufbau reference, instead restricts the excitation space by removing couplings between core and valence configurations.~\cite{Cederbaum_1980,Coriani_2015,Skomorowski_2021}
CVS thereby transforms a core-excited or core-ionized response problem into a ground-state-like problem within a truncated excitation space.
Although these approaches improve numerical robustness by suppressing couplings that may generate complex roots, they modify the response problem and, consequently, the resulting energies.
Excited-state SCF references can serve as starting points for subsequent response calculations, but existing implementations generally rely on TDA or CVS that avoid complex energies. 

In this work, we investigate the origin of complex RPAx energies for ground- and excited-state HF references without imposing these decoupling strategies.
We first consider the dissociation of \ce{H2} in the 6-31G basis set~\cite{Hehre_1972} as a prototypical model.
Its small size permits a direct connection between the HF stationary points, the eigenvalues of the stability matrix, and the appearance of real or complex RPAx energies.
At the same time, the multiconfigurational character of \ce{H2} in the dissociation limit makes it challenging for both restricted HF and non-Hermitian response methods.
Although \ce{H2} has been extensively analyzed in this context,~\cite{Angeli_2013,Scuseria_2013,Cui_2013,Burton_2021,Burton_2022,Marie_2023,Song_2025} RPAx based on its excited-state stationary solutions remains largely unexplored.
We show that these references exhibit instabilities and analyze their origin.

We then assess the practical consequences of going beyond the TDA in spin-flip RPAx and BSE calculations of cyclobutadiene. 
Finally, we examine the effect of removing CVS in state-specific RPAx core-ionization calculations.
Together, these examples demonstrate the conditions under which the complex roots predicted by the stability analysis emerge in realistic calculations, as well as the regimes in which the full, untruncated response equations remain stable and possess an entirely real spectrum.
\section{Hartree-Fock Theory}
\begin{figure*}
  \subfloat[RHF $\sigma_g^2$\label{fig:orbitals_gs_rhf_1.5}]{%
    \includegraphics{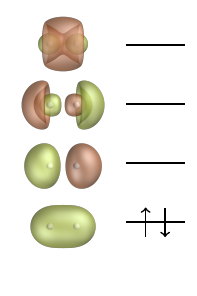}%
  }%
  \hfill%
  \subfloat[Ground-state dissociation\label{fig:GS_RHF_vs_UHF}]{%
    \includegraphics[width=0.49\linewidth]{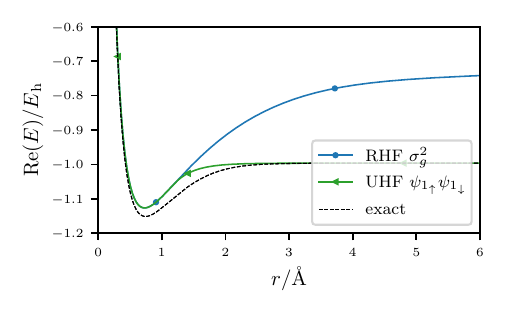}%
  }%
  \hfill%
  \subfloat[UHF $\psi_{1_\upar}\psi_{1_\dwar}$\label{fig:orbitals_gs_uhf_1.5}]{%
    \includegraphics{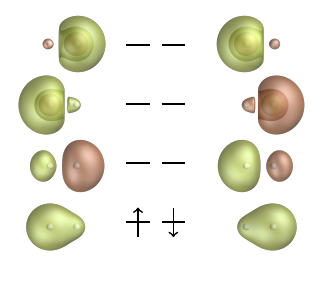}%
  }%
  \caption{
    (a) Illustration of orbitals and occupation for the RHF $\sigma_g^2$ ground state at $r=\SI{1.5}{\angstrom}$.
    (b) Ground-state \ce{H2} dissociation curve in the 6-31G basis set comparing RHF with UHF.
    Exact results are obtained from FCI calculations in the same basis set with \textsc{Quantum Package 2.0}.~\cite{QP2}
    (c) Illustration of orbitals and occupation for the UHF $\psi_{1_\upar} \psi_{1_\dwar}$ ground state at $r=\SI{1.5}{\angstrom}$.
    The orbitals have been visualized with \textsc{pegamoid}~\cite{Pegamoid} at an isovalue $\alpha = 0.05$.
  }
\end{figure*}

The HF method~\cite{Hartree_1928,Fock_1930,Szabo_1996} provides the single-determinant mean-field references used throughout this work.
The HF ground-state energy is variational with respect to orbital rotations
\begin{equation}
	E_\text{HF} = \min_{\ket{\Phi}} \mel{\Phi}{\hat{H}}{\Phi}
	\label{eq:min_HF}
\end{equation}
where $\ket{\Phi}$ is a normalized Slater determinant made of $N$ spin orbitals $\psi_{p}$ obtained from the self-consistent Roothaan-Hall equations.~\cite{Roothaan_1951,Hall_1951}
For ground-state calculations, the orbitals are occupied according to the Aufbau principle, while excited-state calculations require a procedure for targeting a specific stationary solution. 
In this work, we consider the restricted HF (RHF) and unrestricted HF (UHF)\cite{Pople_1954,Fukutome_1981} variants, and use MOM~\cite{Gilbert_2008,Barca_2018,Corzo_2021} to obtain excited-state references.
\subsection{Restricted Hartree-Fock}
For closed-shell RHF, the spatial orbitals are constrained to be identical for the two spin channels, i.e., $\psi_{p_\upar}(\br) = \psi_{p_\dwar}(\br)$.
The resulting determinant is an eigenstate of $\hat{S}_z$ and $\hat{S}^2$, with eigenvalue $0$ in both cases. 
For \ce{H2}, the Aufbau principle therefore leads to the occupation of the bonding $\sigma_g$ orbital by one spin-up ($\upar$) and one spin-down ($\dwar$) electron, as illustrated in Fig.~\ref{fig:orbitals_gs_rhf_1.5}.

We compare the RHF solution with exact full CI (FCI) results in the same 6-31G basis set.~\cite{Helgaker_2000}
As shown in Fig.~\ref{fig:GS_RHF_vs_UHF}, the RHF energy is close to the exact result at short interatomic distances and correctly describes the qualitative shape of the potential energy curve.
The remaining energy difference arises from the lack of electron correlation, and the discrepancy increases substantially toward the dissociation limit.
In particular, a single doubly occupied spatial orbital cannot describe the correct separated-atom limit.
\subsection{Unrestricted Hartree-Fock}
UHF removes the constraint that the spatial orbitals of the two spin channels must be identical, i.e., $\psi_{p_\upar}(\br)\neq\psi_{p_\dwar}(\br)$, and therefore allows the determinant to break spin symmetry. 
Thus, the UHF wavefunction remains an eigenstate of $\hat{S}_z$, but not of $\hat{S}^2$. 
Since UHF minimizes the energy in a variational space that contains the RHF space, it follows that $E_{\text{UHF}} \leq E_{\text{RHF}}$.

At short interatomic distances, the RHF and UHF solutions are equivalent, and the occupied orbitals are both $\sigma_g$.
Beyond the Coulson-Fischer point at approximately $r=\SI{1.2}{\angstrom}$, however, the UHF solution becomes lower in energy than the RHF solution.~\cite{Coulson_1949}
Upon dissociation, UHF correctly approaches the energy of two neutral hydrogen atoms and therefore the exact dissociation limit for \ce{H2}.
As illustrated in Fig.~\ref{fig:orbitals_gs_uhf_1.5}, the spin-up and spin-down electrons localize on opposite hydrogen atoms, thereby breaking both spin and spatial symmetry.
The two possible localization patterns are symmetry-equivalent.
In contrast, RHF cannot describe the separated atoms and overestimates the dissociation energy by more than $\SI{0.2}{\hartree}$.
\subsection{Maximum Overlap Method}
\begin{figure*}
  \raggedright
  \subfloat[RHF $\sigma_u^2$\label{fig:orbitals_sigmau2_rhf_3.0}]{%
    \includegraphics{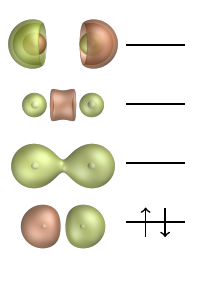}%
  }%
  \hfill
  \subfloat[Doubly excited-state dissociation\label{fig:Ex_RHF_vs_UHF}]{%
    \includegraphics[width=0.48\linewidth]{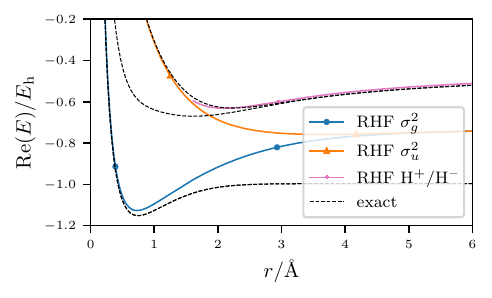}%
  }%
  \hfill
  \subfloat[RHF $\ce{H+/H-}$\label{fig:orbitals_h+h-_rhf_3.0}]{%
    \includegraphics{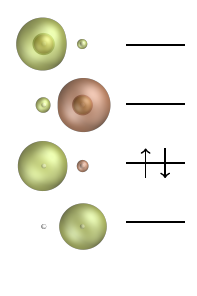}%
  }%
  \caption{
    (b) Dissociation curve of doubly excited \ce{H2} in the 6-31G basis set comparing RHF with UHF.
    Exact results are obtained from FCI calculations in the same basis set with \textsc{Quantum Package 2.0}.~\cite{QP2}
    (a) and (c) Illustration of the orbitals and occupation for a spatially-symmetric $\sigma_u^2$ RHF excited state and a $\ce{H+/H-}$ spatial-symmetry-broken RHF state at an interatomic distance $r=\SI{3.0}{\angstrom}$ calculated using MOM.
    The orbitals have been visualized with \textsc{pegamoid}~\cite{Pegamoid} at an isovalue $\alpha = 0.05$.
  }
\end{figure*}

For excited-state HF calculations,~\cite{Thom_2008,Grazioli_2026,Burton_2021} the variational principle is replaced by a stationary principle, so that the HF energy is required to satisfy
\begin{equation}
	E_\text{HF} = \operatorname*{stat}_{\ket{\Phi}} \mel{\Phi}{\hat{H}}{\Phi}
	\label{eq:stat_HF}
\end{equation}
with a vanishing gradient with respect to orbital rotations.
The challenge is then to converge the self-consistent procedure to the desired excited-state stationary solution, without collapsing to the ground state.
MOM~\cite{Gilbert_2008,Barca_2018,Corzo_2021} addresses this problem by selecting, at each iteration, the orbitals with the largest overlap with a prescribed occupied orbital space.
This procedure can be applied to both RHF and UHF.

For \ce{H2}, we first consider the doubly excited $\sigma_u^2$ configuration shown in Fig.~\ref{fig:orbitals_sigmau2_rhf_3.0}.
A ground-state calculation is used to generate an initial guess, after which the ungerade orbital is targeted and followed using MOM.
At $r=\SI{3.0}{\angstrom}$, the converged occupied $\sigma_u$ orbital is also the lowest-energy orbital, so that the final orbital ordering follows the Aufbau principle.
Nevertheless, the solution remains an excited state because its total energy is higher than that of the $\sigma_g^2$ ground state, as shown in Fig.~\ref{fig:Ex_RHF_vs_UHF}.

A second excited-state solution can be obtained by using localized orbitals as an initial guess and then tracking the corresponding orbital with MOM, as illustrated in Fig.~\ref{fig:orbitals_h+h-_rhf_3.0}.
We obtain those localized orbitals from an SCF calculation on HF$^{8+}$ using the same H basis function on the F atomic center. 
This localized initial guess allows MOM to converge to a second, spatial-symmetry-broken RHF stationary solution.
For $r<\SI{0.3}{\angstrom}$, a UHF solution is obtained with the same energy as the corresponding RHF solution. 
Note that this UHF solution is not visible in Fig.~\ref{fig:Ex_RHF_vs_UHF}, as the corresponding energies lie outside the range shown. 

The symmetric $\sigma_u^2$ RHF solution approaches the ground-state RHF solution in the dissociation limit and becomes degenerate with the $\sigma_g^2$ configuration, as shown in Fig.~\ref{fig:Ex_RHF_vs_UHF}.
In contrast, the symmetry-broken RHF solution approaches a localized $\ce{H+/H-}$ configuration, with both electrons localized on the same hydrogen atom, as illustrated in Fig.~\ref{fig:orbitals_h+h-_rhf_3.0}.
The corresponding exact FCI states have a different character because they are not restricted to a single determinant.
The two states of gerade symmetry arise predominantly from the mixing of the $\sigma_g^2$ and $\sigma_u^2$ configurations, while the third singlet state has predominantly open-shell $\sigma_g\sigma_u$ character.
In the dissociation limit, these three states approach one neutral (covalent) state, corresponding to two neutral hydrogen atoms with one electron on each atom, and two degenerate ionic states corresponding to the symmetric and antisymmetric combinations of the two equivalent $\ce{H+/H-}$ and $\ce{H-/H+}$ configurations.
The excited-state symmetry breaking observed at the RHF level can therefore be viewed as the localization of one of these ionic configurations.
This is analogous to the Coulson-Fischer instability in ground-state Hartree-Fock theory, where a symmetry-broken solution bifurcates from a symmetry-preserved solution.
However, in contrast to the ground-state case, the symmetry-broken excited-state solution remains higher in energy than its symmetry-preserving counterpart.
This distinction will be important when interpreting the complex RPAx roots associated with excited-state references.

\subsection{Stability Analysis}
\label{subsec:stab}
By definition, the HF solution is a stationary point of the energy with respect to orbital rotations, meaning the gradient vanishes.
But what about the Hessian?
\begin{equation} 
  \bH=\mqty(\bA & \bB  \\ \bB^* & \bA^*)
  \label{eq:Hessian}
\end{equation}
In the case of a canonical HF solution with a vanishing gradient, the Hessian is built from the orbital energies $\epsilon_p$ and antisymmetrized two-electron integrals~\cite{Helgaker_2000,Seeger_1977,Cizek_1967}
\begin{subequations} \label{eq:AB}
	\begin{align}
		A_{ia,jb} &= (\epsilon_a - \epsilon_i) \delta_{ij} \delta_{ab} + \abraket{aj}{ib}
		\\
		B_{ia,jb} &= \abraket{ab}{ij}
	\end{align}
\end{subequations}
Here, $i,j$ denote occupied spin orbitals and $a,b$ denote unoccupied spin orbitals, while $\braket{pq}{rs}$ are two-electron integrals in Dirac notation and $\abraket{pq}{rs} = \braket{pq}{rs} - \braket{pq}{sr}$.
The transition index $ia$ thus represents an excitation of an electron from an occupied spin orbital $i$ to an unoccupied spin orbital $a$.
For the ground state, which corresponds to a minimum of the energy, the Hessian must be positive semidefinite.~\cite{Thouless_1960}

Assuming real orbitals and consequently real $\bA$ and $\bB$, the positive definiteness condition can be split into two conditions:
the positive definiteness of $\bA+\bB$, accounting for rotations within the space of real orbitals, and $\bA - \bB$, accounting for rotations towards orbitals with non-zero imaginary parts.~\cite{Seeger_1977}
For a closed-shell RHF reference, spin adaptation combines the corresponding $\upar\upar$ and $\dwar\dwar$ orbital rotations into symmetric and antisymmetric combinations, thereby block-diagonalizing $\bA$ and $\bB$ into independent singlet and triplet sectors.
A negative eigenvalue, and thus violation of the positive definiteness of the Hessian, in the singlet block corresponds to an instability within RHF space.
This means a lower-lying RHF solution exists, and thus we did not converge to a minimum of the RHF energy.
On the other hand, a negative eigenvalue in the triplet block corresponds to an instability toward UHF, meaning a lower-lying UHF solution exists. 
However, the RHF solution remains a minimum if the singlet block is positive definite.

A similar analysis can be performed for UHF by separating spin-conserving and spin-flip orbital rotations. 
The former probes stability within the UHF manifold, whereas the latter probes stability toward generalized HF (GHF).\cite{Seeger_1977,Jimenez-Hoyos_2011}
Since GHF orbitals are general spinors that mix the spin-up and spin-down components, a GHF determinant is, in general, not an eigenfunction of either $\hat{S}_z$ or $\hat{S}^2$.

Looking at the ground state of $\ce{H2}$ in Fig.~\ref{fig:GS_RHF_vs_UHF}, we analyze the stability of the RHF ground-state solution toward UHF.
Before the Coulson-Fischer point, the UHF and RHF solutions are stable, and $\bA+\bB$, as well as $\bA-\bB$ are positive definite.
Then, past the Coulson-Fischer point, the singlet part of $\bA+\bB$ stays positive definite, while the triplet part is not.
The UHF solution is stable all the way.

For excited states, this analysis no longer applies in the same way.
Using MOM, we converge to stationary solutions.
However, an excited-state solution is generally a saddle point rather than a local minimum of the energy.
Consequently, the Hessian is not required to be positive semidefinite and may contain negative eigenvalues associated with orbital rotations toward lower-lying states.
\section{Random-Phase Approximation}
While HF lacks dynamical correlation and thus fails to accurately describe dissociation curves, RPAx incorporates correlation through the coupling between excitation and de-excitation modes.
A related particle-hole formulation is direct RPA (dRPA), which retains only the direct Coulomb contribution and is commonly used to obtain the polarizability entering the screened interaction in $GW$ calculations.~\cite{Berkelbach_2018,Bintrim_2021,Marie_2024a}
Here, we focus on RPAx because its response matrices $\bA$ and $\bB$ are identical to those entering the HF stability analysis (see Sec.~\ref{subsec:stab}).
\subsection{Theory}
\label{subsec:rpax_theory}
The matrices $\bA$ and $\bB$ introduced in Eq.~\eqref{eq:AB} define the particle-hole RPAx response problem, which takes the form of the following Casida-like equation~\cite{Casida_1995,Dreuw_2005,Hesselmann_2011,Chen_2017}
\begin{equation}
	\mqty(\bA & \bB  \\ -\bB^* & -\bA^*) \cdot \mqty(\bX & \bY^* \\ \bY & \bX^*)
	= \mqty(\bX & \bY^* \\ \bY & \bX^*) \cdot \mqty(\bOmega &\bO\\ \bO &-\bOmega^*)
	\label{eq:casida_like}
\end{equation}
where $\bX$ and $\bY$ are matrices whose columns are the vectors $\bX_\nu$ and $\bY_\nu$, representing the excitation and de-excitation amplitudes, respectively.
Throughout this work, we use ``excitation'' to denote a particle-hole transition from the reference determinant to another state, irrespective of its energy relative to the reference.
Thus, transitions to lower-lying states have negative transition energies but are still referred to as excitations from the reference.
We use ``de-excitation'' for the corresponding inverse transition.
The number of particle-hole transitions, $\Ns$, is given by the product of the numbers of occupied and unoccupied orbitals entering $\bA$ and $\bB$. 
The diagonal matrix $\bOmega = \diag(\Omega_1,\ldots,\Omega_\Ns)$ contains one member of each eigenvalue pair of the full RPAx matrix, whose spectrum is given by $\{\Omega_\nu,-\Omega_\nu^*\}$.
The eigenvalues $\Omega_\nu$ correspond to excitation energies relative to the underlying reference and can be either positive or negative.

If the eigenvectors and eigenvalues are real and both $\bA + \bB$ and $\bA - \bB$ are positive definite, the RPAx problem reduces to diagonalizing a Hermitian matrix of half the dimension,~\cite{Marie_2024a} as shown in Eq.~\eqref{eq:red_herm_casida_like_xmy}.
In practice, the RPA equations can be reduced to a non-Hermitian eigenvalue problem of half the size even for excited state references, provided the eigenvectors are real-valued.
For excited-state references, we must recover signs of $\bOmega$ and enforce orthonormalization (see Appendix~\ref{app:excited_casida}).

Once the RPAx equations \eqref{eq:casida_like} are solved for $\bOmega$, the correlation energy can be evaluated using the plasmon formula~\cite{McLachlan_1964,Eshuis_2012}
\begin{equation}
  E_\text{c} = \frac{1}{4} \sum_{\nu=1}^{\Ns} \Omega_\nu - \frac{1}{4} \sum_{ia} A_{ia,ia}
  \label{eq:Ecorr}
\end{equation}
Indices $ia$ denote a pair of occupied and unoccupied orbitals in the HF reference.
This gives the total energy
\begin{equation}
  E_\text{RPAx} = E_{\text{HF}} + E_{\text{c}}
\end{equation}

If the RPAx stability matrix is positive definite, all RPAx excitation energies are real and strictly positive.
If the stability matrix is only positive semidefinite, zero-frequency solutions may additionally occur.~\cite{Ring_1980,Nakada_2016}
When such a zero-frequency solution originates from the spontaneous breaking of a continuous symmetry of the Hamiltonian at the mean-field level, it is referred to as a Nambu-Goldstone mode.~\cite{Nambu_1961,Goldstone_1961,Nakada_2017,Cui_2013}
For example, a spin-symmetry-broken UHF solution of a spin-rotation-invariant electronic Hamiltonian, as can occur for stretched \ce{H2}, selects a particular spin configuration.
Global spin rotations generate energetically equivalent mean-field solutions, and the associated zero-frequency modes appear in RPAx.~\cite{Jimenez-Hoyos_2011,Cui_2013}

However, for excited-state references, negative transition energies $\Omega_\nu$ can arise due to transitions to lower-lying states.
Consequently, a different criterion is required to assign the eigenvectors and eigenvalues of the RPAx matrix to excitations and de-excitations.

To address this, one must sort and orthonormalize the eigenvectors with respect to the indefinite metric~\cite{Scuseria_2013,Song_2025}
\begin{equation}
  \mqty(\bX & \bY^* \\ \bY & \bX^*)^\dagger \cdot \mqty(\bI & \bO \\ \bO & -\bI) \cdot \mqty(\bX & \bY^* \\ \bY & \bX^*) \stackrel{!}{=} \mqty(\bI & \bO \\ \bO & -\bI)
  \label{eq:etanorm}
\end{equation}
This is equivalent to the conditions
\begin{subequations}
\begin{align}
  \bX^\dagger \cdot \bX - \bY^\dagger \cdot \bY & = \bI
  \label{eq:RPA_normalization}
  \\
  \bX^T \cdot \bY - \bY^T \cdot \bX & = \bO
  \label{eq:RPA_orthogonal}
\end{align}
\end{subequations}
The eigenvectors assigned to the excitation sector are normalized to fulfill Eq.~\eqref{eq:RPA_normalization}, while their de-excitation partners have norm minus one, i.e., $(\bY^*)^\dagger \cdot \bY^* - (\bX^*)^\dagger \cdot \bX^* = - \bI$.
Importantly, for excited-state references, the sign of the excitation energy alone does not determine this assignment.
An excitation to a lower-lying state may have a negative transition energy while retaining positive indefinite-metric norm.

If a complete set of RPAx modes can be normalized according to Eqs.~\eqref{eq:RPA_normalization} and \eqref{eq:RPA_orthogonal}, and all excitation-sector eigenvalues are strictly positive, the reference is strictly stable.~\cite{Nakada_2016a}
For each normalized mode, Eq.~\eqref{eq:RPA_normalization} implies
\begin{equation}
  \norm{\bX_\nu}^2-\norm{\bY_\nu}^2=1
\end{equation}
so that the Euclidean norm of the excitation amplitude $\bX_\nu$ exceeds that of the de-excitation amplitude $\bY_\nu$.
This expresses the excitation character of the mode in the chosen RPAx representation, independently of whether its transition energy is positive or negative.

A mode cannot be normalized with respect to the indefinite metric if its norm vanishes, which occurs when $\norm{\bX_\nu}=\norm{\bY_\nu}$.
Such zero-norm modes cannot be represented as conventional normalized RPA modes.
In particular, eigenvectors associated with non-real eigenvalues necessarily have zero indefinite-metric norm.~\cite{Peng_2013}
However, zero norm does not by itself imply a complex eigenvalue, since it may also occur at special degeneracies or for zero-frequency modes.

The vanishing indefinite-metric norm of eigenvectors associated with non-real eigenvalues prevents these modes from being assigned to the excitation or de-excitation sector according to the normalization criterion above.
For the real-valued RPAx matrices considered here, non-real eigenvalues occur either as a purely imaginary pair, $\{\Omega_\nu,-\Omega_\nu\}$, with $\Re(\Omega_\nu)=0$, or as a complex quartet, $\{\Omega_\nu,-\Omega_\nu,\Omega_\nu^*,-\Omega_\nu^*\}$, with $\Re(\Omega_\nu)\neq 0$ and $\Im(\Omega_\nu)\neq 0$.~\cite{Nakada_2016} 
When non-real eigenvalues occur, we adopt the convention of omitting them from the first sum in Eq.~\eqref{eq:Ecorr}.
For a purely imaginary pair, this is equivalent to taking the real part of the plasmon expression while including one of the imaginary solutions in the sum, since the non-real eigenvalues have vanishing real parts.
For a complex quartet, assigning zero net contribution to the quartet is equivalent to selecting the pair $\Omega$ and $-\Omega$, whose contributions cancel.
However, this choice is a convention, since the indefinite-metric normalization does not uniquely determine which members of the quartet should enter the sum. 
A preliminary investigation of these cases through the ring CCD equivalence~\cite{Scuseria_2008} is presented in Appendix~\ref{app:cplxrCCD}.

The first condition in Eq.~\eqref{eq:RPA_normalization} enforces mutual orthonormality of the selected excitation-sector modes with respect to the indefinite metric, whereas Eq.~\eqref{eq:RPA_orthogonal} enforces orthogonality between these modes and their de-excitation partners.
\subsection{Ground-State RPAx Instabilities}
\label{subsec:GS_RPAx}
\begin{figure*}
  \begin{minipage}[b][5cm]{0.45\linewidth}
    \subfloat[Ground state\label{fig:GS_RPAx}]{\includegraphics{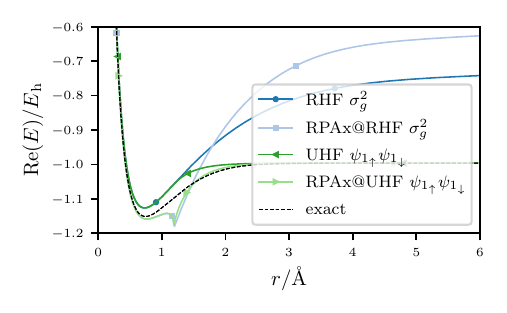}}
  \end{minipage}
  \hfill
  \begin{minipage}[b][5cm]{0.45\linewidth}
    \subfloat[Triplet Excitation\label{fig:h2_triplet_sigma_g2}]{\includegraphics{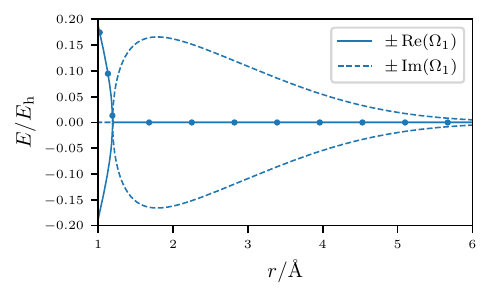}}
  \end{minipage}
  \caption{
    (a) Ground-state \ce{H2} dissociation curve in the 6-31G basis set comparing RPAx on top of a RHF and UHF reference.
    Exact results are obtained from FCI calculations in the same basis set with \textsc{Quantum Package 2.0}.~\cite{QP2}
    (b) Real (solid) and imaginary (dashed) parts of the lowest RPAx triplet excitation energies during the dissociation process of the $\sigma_g^2$ RHF configuration.
    The line marked by dots corresponds to the excitation, whereas the unmarked line is the de-excitation counterpart.
  }
\end{figure*}
In Fig.~\ref{fig:GS_RPAx}, the RPAx energy near equilibrium ($r\approx \SI{0.7}{\angstrom}$) closely matches the FCI result with an absolute error of \SI{6.2}{\milli\hartree}, significantly improving upon HF with an absolute error of \SI{23.9}{\milli\hartree}. 
However, a kink develops in the RPAx energy curve near the Coulson-Fischer point ($\approx \SI{1.19}{\angstrom}$).
As shown in Fig.~\ref{fig:h2_triplet_sigma_g2}, the lowest triplet excitation energy computed on top of the restricted reference vanishes precisely at this point. 
The corresponding excitation and de-excitation eigenvalues of the $\pm\Omega$ pair then become indistinguishable according to the criterion of Scuseria et al.~\cite{Scuseria_2013} given in Eq.~\eqref{eq:etanorm}. 
Consequently, the RPAx matrix becomes singular and loses rank, reflecting the degeneracy induced by the mixing of the excitation with its de-excitation counterpart. 
This singularity is responsible for the kink observed in the RPAx energy curve.

Following the RHF reference beyond the Coulson-Fischer point, the eigenvalue becomes purely imaginary, and the RPAx matrix recovers full rank.
As the dissociation limit is approached, however, the imaginary part progressively vanishes.
The degeneracy of different RHF and UHF solutions is reflected by zero RPAx singlet and triplet excitation energies.
Following the UHF solution beyond the Coulson-Fischer point, the zero eigenvalue stays zero, reflecting the broken spin symmetry.
As the dissociation limit is reached, they reflect the degeneracy of the true ground state.~\cite{Cui_2013}

\subsection{Excited-State RPAx Instabilities}
\label{subsec:ex_rpax}
\begin{figure*}
	\centering
	\includegraphics{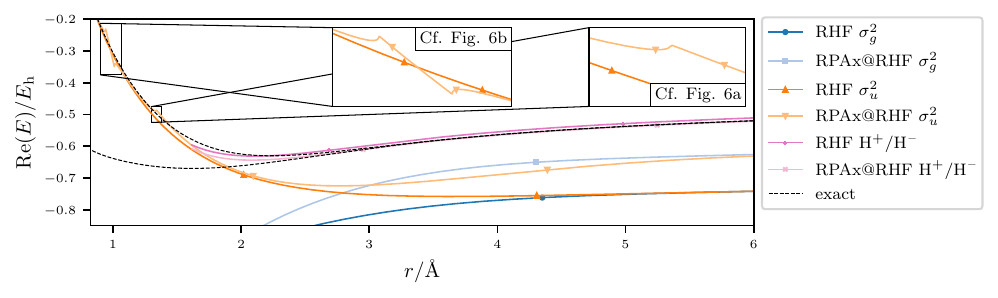}
  \caption{Dissociation curve of \ce{H2} in the 6-31G basis set, comparing RPAx energies for three RHF references: (i) a spatially symmetry-broken solution (converging to \ce{H+/H-} in the dissociation limit), (ii) a $\sigma_u^2$ configuration, and (iii) a $\sigma_g^2$ configuration.
  The latter two converge to the same delocalized dissociation limit.
  Exact results are obtained from FCI calculations in the same basis set using \textsc{Quantum Package 2.0}.~\cite{QP2}
  The left inset highlights a triplet instability near $r = \SI{0.9}{\angstrom}$, while the right inset shows a singlet instability of the spatially symmetric reference near $r = \SI{1.2}{\angstrom}$.
	}
	\label{fig:Ex_RPAx}
\end{figure*}
\begin{figure*}
  \centering
  \subfloat[Singlet Excitation\label{fig:h2_singlet2_sigma_u2}]{\includegraphics{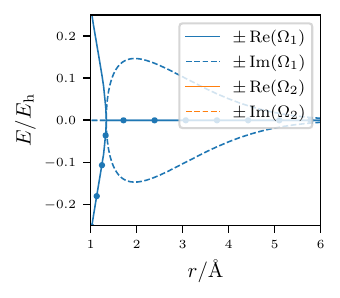}}
  \hfill
  \subfloat[Triplet Excitations\label{fig:h2_triplet_sigma_u2}]{\includegraphics{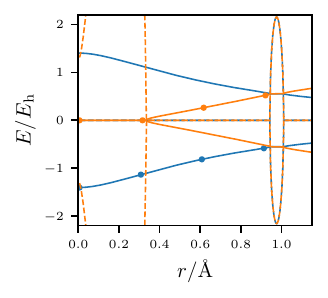}}
  \hfill
  \subfloat[Singlet Excitations\label{fig:h2_singlet1_sigma_u2}]{\includegraphics{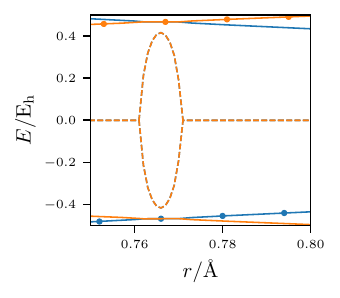}}
  \caption{Real (solid) and imaginary (dashed) parts of the lowest RPAx excitation energies during the dissociation process of the $\sigma_u^2$ RHF configuration. The line marked by dots corresponds to the excitation, whereas the unmarked line is the de-excitation counterpart. 
  The legend in panel (a) applies to all three panels.
}
\end{figure*}
Following the $\sigma_u^2$ RHF excited state (see Fig.~\ref{fig:Ex_RPAx}), we observe a similar behavior to the ground state around $r = \SI{1.2}{\angstrom}$.
One excitation energy vanishes, mixing with its own de-excitation, as shown in Fig.~\ref{fig:h2_singlet2_sigma_u2}.
At the point where it reaches exactly zero, the excitation becomes indistinguishable from its de-excitation counterpart, and an imaginary contribution to the excitation energy arises.
In the dissociation curve, we observe an unphysical kink, as seen in the right inset of Fig.~\ref{fig:Ex_RPAx}.
This vanishing excitation--de-excitation pair arises from the singlet block rather than the triplet block, as in the ground-state instability.
This indicates an instability of the RHF solution within the RHF space.
The corresponding Hessian eigenvector identifies a spatial-symmetry-breaking direction.
We locate a spatially symmetry-broken stationary solution only from approximately $r=\SI{1.6}{\angstrom}$ onward (pink curve) and were unable to converge this solution near $r=\SI{1.2}{\angstrom}$.
Therefore, although the response instability at $r=\SI{1.2}{\angstrom}$ is consistent with spatial symmetry breaking, our calculations do not establish numerically that the located symmetry-broken branch bifurcates exactly at this point.
Another difference from the ground-state RPAx instability is that the excitation-sector mode involved here has negative energy before reaching zero.
Such behavior is possible for an excited-state reference, which supports transitions to lower-lying states, but not for a stable ground-state reference.
Regarding the RPAx eigenvalues of the symmetry-broken solution towards the dissociation limit, we do not find any zero excitation energies, but only real, negative and positive RPAx excitation energies in the singlet and triplet manifold.

A triplet instability of the $\sigma_u^2$ RHF solution, similar to the one of the ground state, is observed at small $r$, as shown in Fig.~\ref{fig:h2_triplet_sigma_u2}.
This is outside of the scope of Fig.~\ref{fig:Ex_RPAx}.
Here, a UHF solution emerges with identical energy to the RHF solution.
Consequently, a triplet transition appears with a vanishing real excitation energy but a non-zero imaginary part.
As the distance $r$ increases, this instability disappears, and the first triplet transition becomes real and non-zero at around $r = \SI{0.3}{\angstrom}$.

Another type of instability, unique to excited states, arises from the degeneracy between a positive excitation and a positive de-excitation (the counterpart to a negative excitation).
While the previously discussed instabilities are linked to spin- or spatial-symmetry breaking, this type is not.
These instabilities only occur when negative excitations are present.
In the $\ce{H2}$ dissociation curve of the $\sigma_u^2$ RHF solution in Fig.~\ref{fig:Ex_RPAx}, we find two examples of such instabilities.
The first is a triplet instability between approximately $r = \SI{0.9}{\angstrom}$ and $r = \SI{1.0}{\angstrom}$ as seen in Fig.~\ref{fig:h2_triplet_sigma_u2}. 
Here, the first two triplet excitation energies mix and become degenerate. 
This causes a kink in the dissociation curve, similar to the other instabilities, as seen in the left inset of Fig.~\ref{fig:Ex_RPAx}.
Subsequently, the excitations become distinguishable again, causing another kink in the curve.
This corresponds to the left inset of Fig.~\ref{fig:Ex_RPAx}.
The second case involves the mixing of a singlet excitation with a singlet de-excitation around $r = \SI{0.75}{\angstrom}$, as shown in Fig.~\ref{fig:h2_singlet1_sigma_u2}.
The qualitative behavior mirrors that of the previously discussed triplet instability.
We also observe an unsmooth section in the dissociation curve corresponding to this instability.
The energy curve exhibits two kinks qualitatively similar to those shown in the left inset of Fig.~\ref{fig:Ex_RPAx}, which arise from the triplet-mode collision discussed above.
This feature, however, falls outside the range of the plot in  Fig.~\ref{fig:Ex_RPAx}.

\section{Tamm-Dancoff approximation}
\label{sec:bse_tda_example}
The analysis of the previous sections demonstrated that the instabilities of full RPAx originate from the coupling between excitations and de-excitations through the $\bB$ matrix.
The TDA removes this coupling by neglecting $\bB$, thereby reducing the non-Hermitian RPA eigenvalue problem to the Hermitian CIS problem.
Consequently, the mechanism responsible for the appearance of complex excitation energies cannot occur within the TDA.
This immediately raises the question of how much physics is sacrificed when this approximation is employed in practical calculations.

Spin-flip calculations provide an ideal framework to address this question.
Indeed, spin-flip calculations are commonly performed within the TDA, whereas full RPAx (i.e., beyond-TDA) spin-flip calculations are rarely considered.~\cite{Blase_2018,Blase_2020,Monino_2021}
Here, we investigate the consequences of relaxing the TDA in both RPAx and BSE calculations.
It has been suggested that the use of the TDA is advantageous for spin-flip calculations because it removes instabilities in the response equations, ensuring that excitation energies and energy gradients remain finite even near state crossings.~\cite{Li_2012}
However, when the TDA is released and the full RPA is employed, the excitation energies become closely connected to the orbital stability conditions and correctly reproduce the approach to electronic instabilities.
This comes at the cost of unphysical curvature of the excited-state energies as they approach crossings with the reference state, originating from the mathematical structure of the RPA eigenvalue equations.~\cite{Russ_2004}

As an example, we consider the rectangular $D_\text{2h}$ and square $D_\text{4h}$ geometries of cyclobutadiene reported in Ref.~\citenum{Levchenko_2004}, both optimized at the CCSD(T)/cc-pVTZ level of theory (see also the \SupMat). The former is the singlet ground-state geometry, whereas the latter is the triplet ground-state geometry.

The singlet ground state of the square geometry exhibits pronounced multireference character and is therefore poorly described by single-reference methods.
In contrast, the high-spin ($m_S=1$) triplet state is well described by a single Slater determinant.~\cite{Levchenko_2004,Monino_2021,Casanova_2020}
Spin-flip methods therefore use this triplet state as a reference and recover the singlet manifold through spin-flip excitations.~\cite{Krylov_2001a,Sears_2003,Casanova_2012,Casanova_2020}
This strategy provides access to both singly and doubly excited singlet states while avoiding a multireference reference wavefunction.

However, the triplet reference is itself an excited state.
Consequently, one spin-flip transition necessarily corresponds to a negative-energy excitation towards the singlet ground state.
This is precisely the situation analyzed throughout the previous sections for excited-state references.
Within full RPAx or BSE, this negative excitation can couple to positive excitations through the $\bB$ matrix, potentially leading to the instabilities discussed above.
Within the TDA, this coupling is absent, which explains why spin-flip calculations are almost exclusively performed within this approximation.

The formalism presented in Appendix~\ref{app:excited_casida} allows us to solve the full RPAx and BSE equations without invoking the TDA.
We can therefore directly quantify the influence of the excitation-de-excitation coupling that is neglected within the TDA.
Since the BSE generally provides more accurate neutral excitation energies than RPAx, we consider both methods in the following.
The BSE equations possess the same mathematical structure as the RPA equations in Eq.~\eqref{eq:casida_like}, differing only in the definition of the $\bA$ and $\bB$ matrices.~\cite{Blase_2018, Monino_2021}
Consequently, exactly the same solution strategy can be applied once these matrices have been constructed.

All calculations have been performed in the cc-pVTZ~\cite{Dunning_1989} basis set with \textsc{QuAcK}~\cite{QuAcK} on top of a UHF high-spin triplet reference determinant.
For BSE, we performed a dRPA calculation on the spin-conserved excitation manifold without TDA to obtain the screening, followed by $G_0W_0$ quasiparticle energies solved via the Newton-Raphson procedure.
\begin{table}
  \caption{Comparison of spin-flip excitation energies for cyclobutadiene in the rectangular $D_{2h}$ geometry.
           Excitation energies are given with respect to the singlet $X\,^{1}A_g$ ground state and reported in $\si{\eV}$.
           The TDA effect is quantified as the difference $E_{\text{TDA}}-E_{\text{full}}$, where positive values indicate that the TDA increases the excitation energy.}
  \label{tab:cbd_d2h}
  \begin{ruledtabular}
    \begin{tabular}{l
                  S[table-format=+2.3]
                  S[table-format=+2.3]
                  S[table-format=+2.3]}
      Method
      & {$1\,^{3}B_{1g}$}
      & {$1\,^{1}B_{1g}$}
      & {$2\,^{1}A_g$} \\
      \hline
      SF-TDA-RPAx\fnm[1]
        &1.523 &3.845 &5.508 \\
      \hline
      SF-TDA-BSE
        &1.436 &2.708 &4.540 \\
      SF-BSE
        &1.678 &2.966 &4.743 \\
      $E_{\text{TDA}}-E_{\text{full}}$
        &-0.242 &-0.258 &-0.203 \\
      \hline
      EOM-SF-CCSD(fT)\fnm[2]
        &1.516 &3.260 &4.205
    \end{tabular}
  \end{ruledtabular}
  \fnt[1]{See also Ref.~\citenum{Casanova_2020}.}
  \fnt[2]{Values from Ref.~\citenum{Manohar_2008}.}
\end{table}
\begin{table}
  \caption{Comparison of spin-flip excitation energies for cyclobutadiene in the square $D_{4h}$ geometry.
           Excitation energies are given with respect to the singlet $X\,^{1}B_{1g}$ ground state and reported in $\si{\eV}$.
           The TDA effect is quantified as the difference $E_{\text{TDA}}-E_{\text{full}}$, where positive values indicate that the TDA increases the excitation energy.}
  \label{tab:cbd_d4h}
  \begin{ruledtabular}
    \begin{tabular}{l
                  S[table-format=+1.3]
                  S[table-format=+1.3]
                  S[table-format=+1.3]}
      Method
      & {$1\,^{3}A_{2g}$}
      & {$2\,^{1}A_{1g}$}
      & {$1\,^{1}B_{2g}$} \\
      \hline
      SF-TDA-RPAx\fnm[1]
      &0.318 &3.131 &2.657 \\
      \hline
      SF-TDA-BSE
        &-0.084 &1.204 &1.495 \\
      SF-BSE
        &0.224 &1.528 &1.959 \\
      $E_{\text{TDA}}-E_{\text{full}}$
        &-0.308 &-0.324 &-0.464 \\
      \hline
      EOM-SF-CCSD(fT)\fnm[2]
        &0.163 &1.530 &1.921
    \end{tabular}
  \end{ruledtabular}
  \fnt[1]{See also Ref.~\citenum{Casanova_2020}.}
  \fnt[2]{Values from Ref.~\citenum{Manohar_2008}.}
\end{table}
Tables~\ref{tab:cbd_d2h} and \ref{tab:cbd_d4h} summarize the SF-TDA-RPAx results and the SF-BSE calculations performed with and without the TDA, together with accurate theoretical reference values. The full SF-RPAx results are not included in the tables because their transition energies are complex; their real and imaginary parts are reported in the \SupMat.
The comparison allows us to address two questions.
First, does going beyond the TDA lead to instabilities in practical spin-flip calculations?
Second, if the calculation produces only real energies, does retaining the excitation-de-excitation coupling improve the predicted excitation energies?

In our cyclobutadiene calculations, removing the TDA has distinct consequences for RPAx and BSE. For SF-RPAx, the full response equations give rise to complex transition energies.
In particular, the negative-energy transition from the high-spin triplet reference to the $X\,^{1}A_g$ singlet ground state mixes with positive-energy excitations through the $\bB$ matrix, resulting in complex eigenvalues with an imaginary part of approximately $\pm 2~\mathrm{eV}$  for both geometries. 
Consequently, the real parts of these eigenvalues do not represent physical excitation energies.
They are reported in the \SupMat.

For SF-BSE, in contrast, no such instability occurs.
All transition energies remain real for both geometries, with deviations from the SF-TDA-BSE results of $\SIrange{0.20}{0.46}{\eV}$.
This different behavior may be related to the screened electron-hole interaction employed in BSE, which provides a different description of the coupling between excitations and de-excitations.
In the exact limit, the excitation energies of a Hermitian electronic system are real.
Approaching this limit may therefore be expected to mitigate the occurrence of unphysical complex excitation energies.
The absence of instabilities in SF-BSE for the systems considered here is consistent with this interpretation, although it does not imply that BSE is generally free of instabilities.
Comparing the SF-BSE results with EOM-SF-CCSD(fT), we find no evidence that going beyond the TDA improves the excitation energies relative to SF-TDA-BSE.

\section{Core-valence separation}
In the previous section, we have seen examples where complex energies arise when applying RPAx on an excited-state reference.
As we will explore in the following, for core-hole states, total RPAx energies often stay real.
We thus consider core ionization of \ce{H2O}, \ce{CO}, \ce{BF}, \ce{N2}, and \ce{CO2} using the aug-cc-pCVTZ basis set.~\cite{Dunning_1989,Kendall_1992,Woon_1995,Prascher_2010}
The geometries are taken from Ref.~\citenum{Marie_2026} and, where applicable, from the QUEST database,~\cite{Loos_2020,Veril_2021,Loos_2025} which reports CC3/aug-cc-pVTZ ground-state geometries optimized with CFOUR.~\cite{Matthews_2020a} All geometries used in this work are provided in the \SupMat.
For each system, we perform state-specific RPAx calculations for both the neutral molecule and the core-ionized cation. The $\Delta$RPAx core-ionization energy is then obtained as the difference between their total RPAx energies.

Here, CVS is implemented by excluding all excitations involving unoccupied core orbitals in the construction of $\bA$ and $\bB$ for the state-specific calculation of the cation, and thus the core-hole state.
The RPAx matrix in Eq.~\eqref{eq:casida_like} is then constructed using these reduced response matrices and solved in the same way as the full-space RPAx problem.
In the state-specific $\Delta$CVS-FCI calculations reported in Ref.~\citenum{Marie_2026}, the determinant space is restricted to configurations in which the targeted core orbital remains unoccupied, i.e., configurations that retain the core hole.
CVS schemes have been observed to affect core IPs as they remove spurious couplings between core-ionized states and the valence continuum.~\cite{Liu_2019}
These couplings can lead to unphysical shifts in ionization energies or artificial mixing of states, which CVS mitigates by restricting the excitation space.
The purpose of these examples is to assess the impact of CVS on $\Delta$RPAx core-ionization energies and to determine whether the CVS approximation removes any RPAx instabilities encountered in the corresponding non-CVS calculations.

For \ce{H2O}, CVS shifts the IP, associated with the removal of one 1s electron of the oxygen, by \SI{0.38}{eV}.
Meanwhile, the difference between $\Delta$CVS-RPAx and $\Delta$CVS-FCI is \SI{0.93}{eV}.
This shows that, for some systems, the numerical effect of the CVS approximation is in the same order of magnitude as the residual $\Delta$RPAx error relative to the corresponding $\Delta$CVS-FCI result.
Compared to $\Delta$UHF, which deviates from $\Delta$CVS-FCI by \SI{0.86}{eV}, $\Delta$RPAx exhibits a smaller deviation for this system.
For \ce{CO}, the effect of CVS is similar, with a shift of ~\SI{0.39}{eV}.
Here, the difference between $\Delta$CVS-RPAx and $\Delta$CVS-FCI (\SI{1.14}{eV}) is larger than the CVS shift, but comparable to the error of $\Delta$UHF relative to $\Delta$CVS-FCI (\SI{1.17}{eV}).

For \ce{H2O}, we performed a stability analysis using the same $\bA$ and $\bB$ matrices as those employed in the CVS-RPAx calculation, which enter the Hessian according to Eq.~\eqref{eq:Hessian}.
This Hessian corresponds to the orbital Hessian projected onto the orbital-rotation space retained in the CVS-RPAx calculation.
Importantly, the mathematical connection between the orbital Hessian and the RPAx matrix, when constructed from identical $\bA$ and $\bB$ matrices, is independent of the numerical values of these matrices.
Consequently, if the reference is stable with respect to all orbital rotations in the CVS response space, the corresponding CVS-RPAx eigenvalues are guaranteed to be real-valued.
Conversely, an instability of the reference within this subspace can lead to complex CVS-RPAx eigenvalues.
Thus, the CVS truncation itself does not guarantee a real-valued RPAx spectrum.
Rather, the spectrum inherits the stability properties of the reference within the retained orbital-rotation space.
For \ce{H2O}, this analysis confirms that the MOM reference is stable within the CVS response space.

For the core ionization of fluorine in \ce{BF}, the stability analysis reveals two distinct MOM-UHF solutions for the core-ionized cation.
The first row for \ce{BF} in Table~\ref{tab:core_ip_delta} reports the solution obtained by using the ground-state orbitals of \ce{BF+}, with the occupations appropriately modified, as the initial guess for the MOM calculation.
This procedure yields the same $\Delta$UHF value as that reported in Ref.~\citenum{Marie_2026}.
However, both CVS-RPAx and RPAx calculations performed on this cationic reference exhibit complex excitation energies in the spin-conserved and spin-flip manifolds, indicating the existence of a lower-lying HF solution accessible through orbital rotations within the CVS space.

To locate this solution, we compute the Hessian in the CVS space and use the eigenvector associated with its lowest negative eigenvalue to generate an orbital-rotated guess, while retaining the same orbital occupations.
A subsequent MOM-UHF calculation then converges to the second solution reported for \ce{BF} in Table~\ref{tab:core_ip_delta}.
This solution is stable with respect to spin-conserved UHF orbital rotations within the CVS space: the corresponding block of the CVS Hessian is positive semidefinite, and the CVS-RPAx excitation energies are therefore real in this sector.
The spin-conserved RPAx excitation energies are likewise real.

In contrast, both RPAx and CVS-RPAx still yield complex excitation energies in the spin-flip sector.
These are associated with negative Hessian eigenvalues, revealing that the CVS Hessian is not positive semidefinite with respect to spin-flip rotations.
The resulting instability is therefore toward a GHF solution, implying the existence of a lower-lying GHF state within the CVS-restricted orbital space.
A detailed investigation of these GHF solutions is beyond the scope of the present work, and we do not pursue them further here.

\begin{figure}
  \subfloat[Symmetry broken occupied\label{fig:CO2_broken_alpha}]{%
    \includegraphics[width=0.49\linewidth]{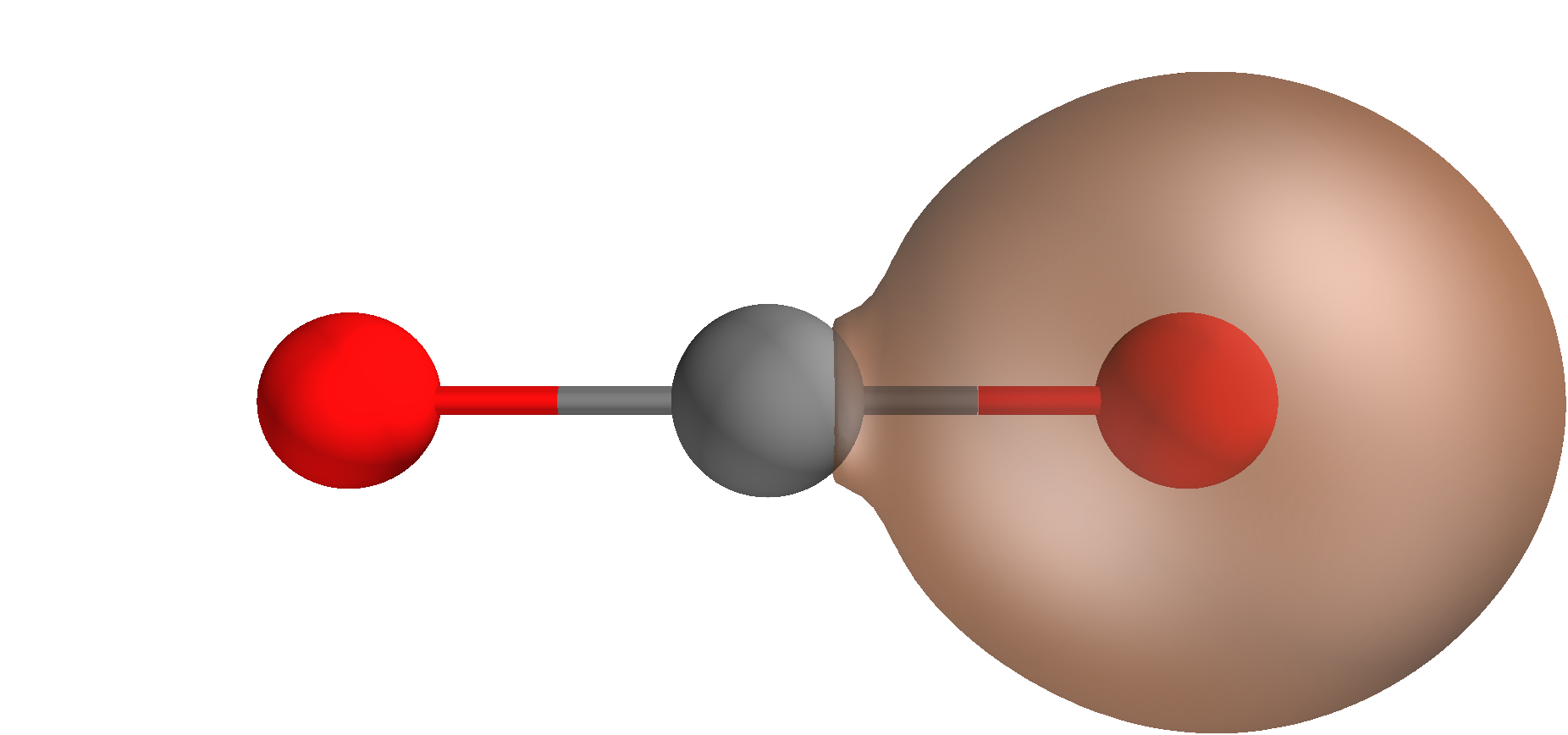}%
  }%
  \hfill%
  \subfloat[Symmetry broken unoccupied\label{fig:CO2_broken_beta}]{%
    \includegraphics[width=0.49\linewidth]{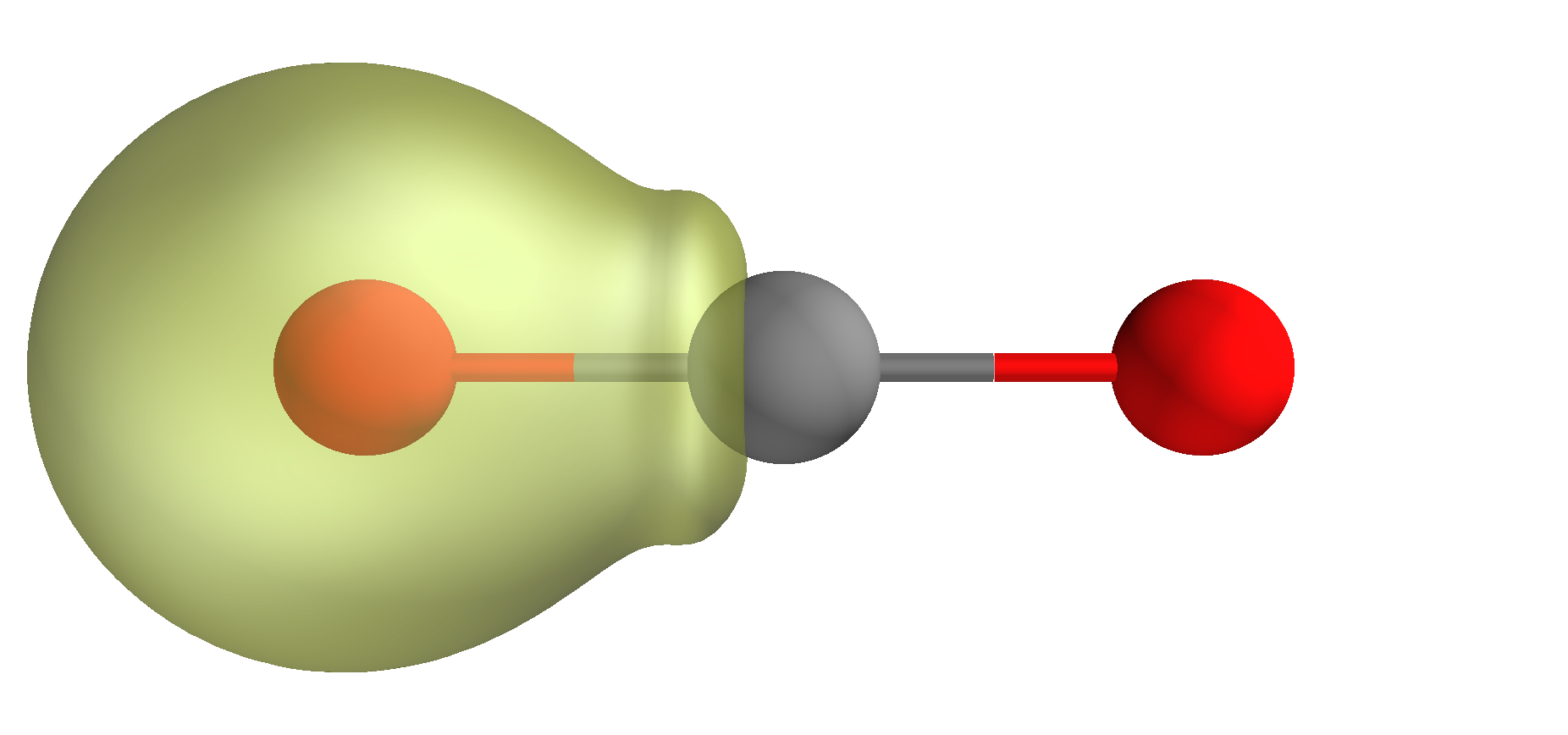}%
  }\\
  \subfloat[Symmetric occupied\label{fig:CO2_symmetric_alpha}]{%
    \includegraphics[width=0.49\linewidth]{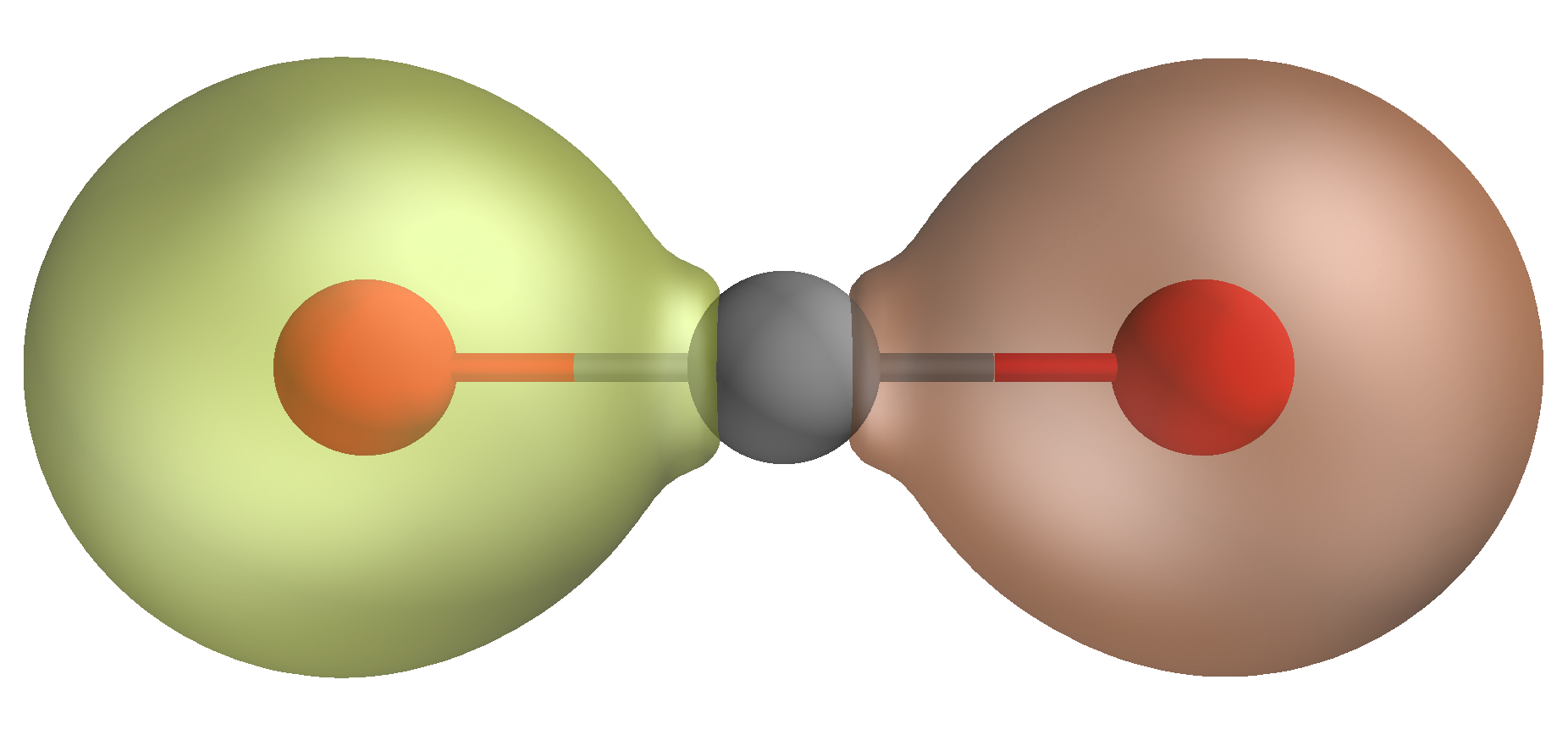}%
  }%
  \hfill%
  \subfloat[Symmetric unoccupied\label{fig:CO2_symmetric_beta}]{%
    \includegraphics[width=0.49\linewidth]{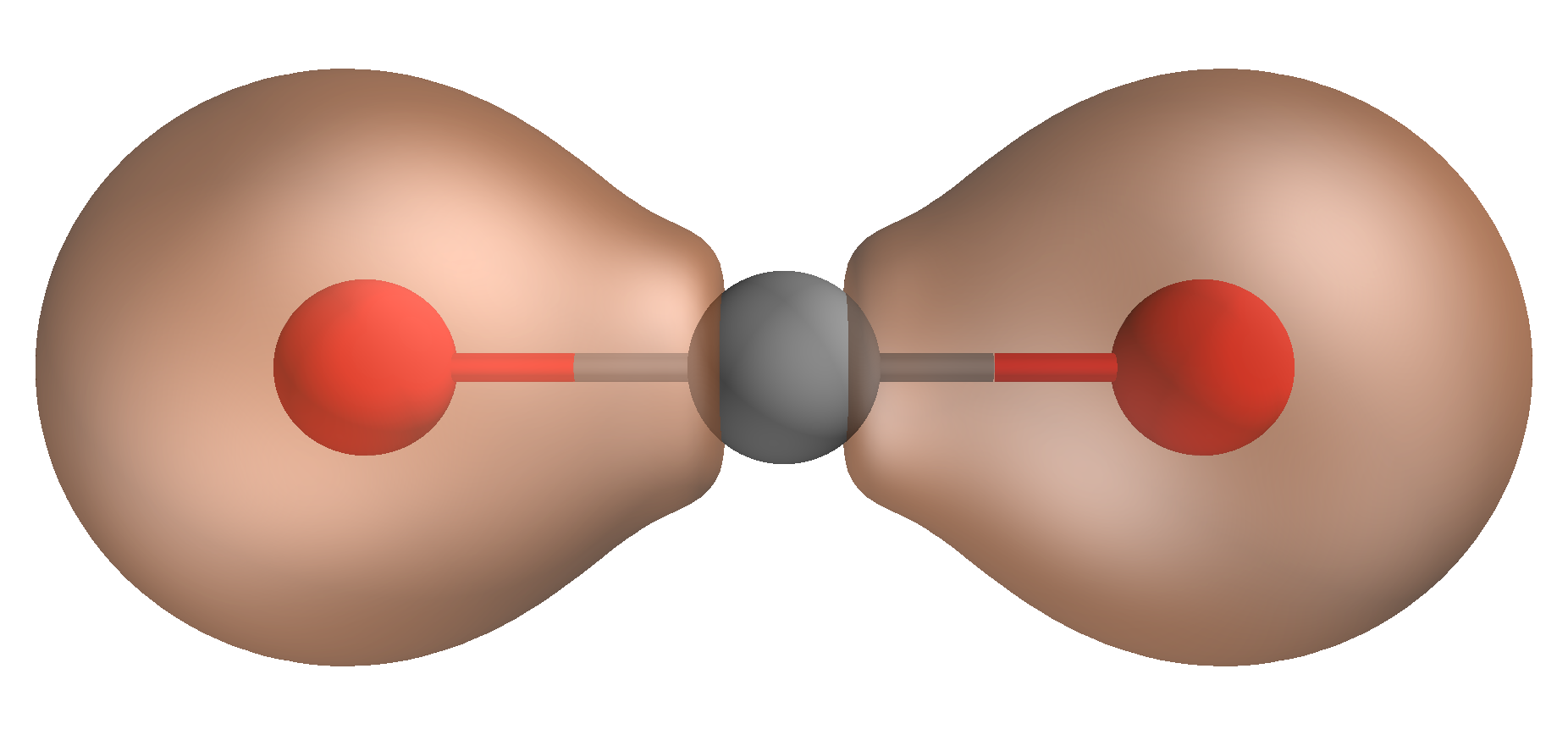}%
  }%
  \caption{
    Occupied and unoccupied core orbitals of oxygen in \ce{CO2+} obtained from MOM-UHF calculations with a symmetry-broken and a symmetric guess.
    The orbitals have been visualized with \textsc{pegamoid}~\cite{Pegamoid} at an isovalue $\alpha = 0.0002$.
  }
  \label{fig:CO2_core_orbitals}
\end{figure}

We also report results for two systems that have degenerate core orbitals: \ce{N2} and \ce{CO2}.
For such systems, breaking spatial symmetry to localize the core orbitals significantly improves the mean-field results compared to experiment.~\cite{Bagus_1972,Cederbaum_1977,Marie_2026}
Here, we compare two approaches for obtaining the mean-field reference for the cation.
First, we perform a RHF calculation for the ground state and generate a guess for a MOM-UHF calculation for the cation by removing one electron from the lowest-lying core orbital.
Second, we perform a RHF for the ground state and then mix the 1$\sigma_g$ and 1$\sigma_u$ core orbitals to localize the core orbitals on the corresponding atoms.
Afterward, we perform a MOM-UHF calculation using a guess in which an electron is removed from one of the core orbitals.
The converged orbitals from the MOM-UHF calculation for the cation in both cases are visualized for \ce{CO2} in Fig.~\ref{fig:CO2_core_orbitals}.

In Table~\ref{tab:core_ip_delta}, the first line for \ce{N2} and \ce{CO2} presents the results for the symmetry-broken solution, whereas the second line presents the results for the symmetric reference.
As we previously observed for the excited state of \ce{H2} in Sec.~\ref{subsec:ex_rpax}, in core-ionized \ce{N2}, RPAx produces unphysical complex transition energies in the spin-flip excitation manifold when using the symmetry-preserved reference.
Applying the CVS approximation does not remove this instability, and complex energies are obtained in the CVS-RPAx calculation as well.
Using the symmetry-broken solution as a reference, all energies remain real.
For \ce{CO2}, the symmetry-preserved solution of the cation remains sufficiently stable to yield real RPAx excitation energies.
For the symmetry-preserved cationic references of both \ce{N2} and \ce{CO2}, the CVS approximation shifts the RPAx core IP by more than \SI{5}{\eV}.
By contrast, the corresponding shifts for the localized core-hole references are only \SI{0.51}{\eV} for \ce{N2} and \SI{0.38}{\eV} for \ce{CO2}.
The large shifts obtained with the symmetry-preserved references can be attributed to the equivalent core centers and the resulting delocalized core orbitals, for which the CVS truncation has a more pronounced effect on the response space.

In summary, the present results show that $\Delta$RPAx can provide reasonable core-ionization energies for realistic systems even without a CVS approximation.
However, neither the full-space nor the CVS-RPAx treatment is guaranteed to yield only real excitation energies when the underlying core-hole reference is unstable.

For the localized or nondegenerate core-hole references considered here, CVS shifts the core IPs by \SIrange{0.38}{0.51}{\eV}. 
$\Delta$RPAx and $\Delta$CVS-RPAx yield results comparable to those of $\Delta$UHF when compared to $\Delta$CVS-FCI.
However, for systems with degenerate core orbitals, such as \ce{N2}, a symmetry-broken, localized core-hole reference is essential to avoid complex RPAx energies.
In this case, symmetry breaking is also essential to obtain reasonable $\Delta$SCF core-ionization energies.
More generally, among the systems investigated here, we did not identify any case in which the CVS approximation removes a complex-energy instability that is present in the corresponding non-CVS RPAx calculation.
Thus, in the present set of core-ionization calculations, CVS does not appear to cure the underlying RPAx instabilities.
Overall, these results suggest that the stability and nature of the core-hole reference are more important for the robustness of $\Delta$RPAx than the CVS truncation itself.
A stability analysis within the CVS space has been useful for finding UHF solutions of core-ionized states.

We note that analogous instabilities may also occur in dRPA when the instability occurs in the spin-conserving block, so the occurrence of complex roots is not specific to the exchange contribution in RPAx.
\begin{table} 
  \caption{Effect of the CVS approximation on core-ionization energies (\si{eV}).
           The asterisk indicates the atom from which the core electron is removed.
           For \ce{N2} and \ce{CO2}, the first row represents the symmetry-broken localized core-hole reference, while the second row represents the symmetry-preserved reference.}
  \label{tab:core_ip_delta}
  \begin{ruledtabular}
    \begin{tabular}{lcccc}
      Molecule               & $\Delta$UHF     & $\Delta$CVS-RPAx     & $\Delta$RPAx          & $\Delta$CVS-FCI\fnm[1] \\
      \hline
      \ce{H2O^*}             & 539.01          & 540.80               & 540.42                & 539.87                 \\ 
      \ce{CO^*}              & 541.30          & 543.61               & 543.22                & 542.47                 \\
      \hline
      \ce{BF^*}              & 694.32          & 696.66\rlap{\fnm[2]} & 696.24\rlap{\fnm[2]}  & 695.43                 \\
      \ce{BF^*}              & 694.20          & 696.53\rlap{\fnm[3]} & 696.11\rlap{\fnm[3]}  &                        \\
      \hline
      \ce{N^*2}              & 409.42          & 412.45               & 411.94                & 409.88                 \\ 
      \ce{N^*2} (sym.)       & 419.25          & 419.10\rlap{\fnm[3]} & 413.15\rlap{\fnm[3]}  &                        \\ 
      \hline
      \ce{CO^*2}             & 540.53          & 541.13               & 540.75                & 541.24                 \\
      \ce{CO^*2} (sym.)      & 552.40          & 553.86               & 548.38                &                        \\
    \end{tabular}
  \end{ruledtabular}
  \fnt[1]{Values from Ref.~\citenum{Marie_2026}.}
  \fnt[2]{RPAx eigenvalues with nonzero imaginary component in spin-conserved and spin-flip sectors.}
  \fnt[3]{RPAx eigenvalues with nonzero imaginary component in spin-flip sector.}
\end{table}

\section{Conclusion}\label{sec:conclusion}
Using the example of \ce{H2}, we have demonstrated that RPAx instabilities can arise due to symmetry breaking, manifesting as complex excitation energies.
For the ground state, this occurs through spin-symmetry breaking towards a UHF solution, while for excited states, spatial-symmetry breaking towards configurations such as \ce{H+/H-} leads to similar behavior.
In both cases, the instability is associated with the vanishing of the real part of an excitation energy, signaling a transition to a symmetry-broken solution.
Furthermore, for excited-state references, an additional type of instability arises from the mixing of positive excitations with the positive counterpart of transitions to lower-lying states.

These findings underscore the importance of understanding and properly handling RPAx instabilities, particularly for applications to excited states and for methods such as $GW$ and BSE, which rely on RPA-based screening.
Our results reveal a stark contrast between RPAx and BSE.
While full SF-RPAx yields complex excitation energies without the TDA, SF-BSE remains stable, with only modest energy shifts of $\SIrange{0.2}{0.5}{\eV}$.
This demonstrates that the TDA is not merely a numerical stabilization but a physical approximation whose removal has drastically different consequences for RPAx (instability) and BSE (small corrections).

A different picture emerges for core ionization.
For sufficiently isolated core-hole states, $\Delta$RPAx can yield entirely real energies provided that a stable core-hole reference is employed.
For the localized or nondegenerate core-hole references considered here, CVS shifts the resulting core IPs by \SIrange{0.38}{0.51}{\eV}; for the symmetry-preserved \ce{N2} and \ce{CO2} references, the shifts exceed \SI{5}{\eV}.
Nevertheless, CVS does not by itself eliminate RPAx instabilities.
Rather, the stability of the underlying core-hole reference appears to be the decisive factor controlling the robustness of $\Delta$RPAx.
This is particularly evident in systems with symmetry-equivalent or degenerate core orbitals, such as \ce{N2}, where symmetry breaking is required to obtain a stable reference and avoid complex energies.

Although the present study is restricted to \ce{H2} and a small set of molecular test cases, the mechanisms identified here are sufficiently general that analogous instabilities can be expected in larger systems, particularly when multireference character or competing symmetry-broken solutions are present.
Core-hole states appear comparatively less prone to such difficulties when the targeted hole is well separated from other electronic states, but the stability of the reference remains essential.
More broadly, these results suggest that stability analysis should accompany state-specific RPAx calculations whenever symmetry breaking or near-degeneracies may be relevant.

\appendix
\section{Practical Solution of the RPA Equations}\label{app:excited_casida}
Given real-valued symmetric matrices $\bA$ and $\bB$, and real RPAx eigenvectors and eigenvalues, we can rewrite the RPAx equations \eqref{eq:casida_like} by exploiting their block structure.
Rather than solving for the full eigenvectors and spectrum, one can solve directly for $\bX+\bY$, $\bX-\bY$, and $\bOmega$.
By adding the two matrix equations from the first line, $\bA\cdot\bX+\bB\cdot\bY=\bX\cdot\bOmega$ and $\bA\cdot\bY+\bB\cdot\bX=-\bY\cdot\bOmega$, we obtain
\begin{equation}
  (\bA+\bB)\cdot(\bX+\bY)=(\bX-\bY)\cdot\bOmega
  \label{eq:red_casida_like_circle}
\end{equation}
Similarly, by subtracting the two equations emerging from the second line, $-\bB\cdot\bX-\bA\cdot\bY=\bY\cdot\bOmega$ and $-\bB\cdot\bY-\bA\cdot\bX=-\bX\cdot\bOmega$, we obtain
\begin{equation}
  (\bA-\bB)\cdot(\bX-\bY)=(\bX+\bY)\cdot\bOmega
  \label{eq:red_casida_like_diamond}
\end{equation}
Solving Eq.~\eqref{eq:red_casida_like_diamond} for $\bX+\bY$ and substituting the result into Eq.~\eqref{eq:red_casida_like_circle} yields~\cite{Furche_2001,Scott_2023}
\begin{subequations}
  \begin{align}
    (\bA-\bB)\cdot(\bX-\bY)\cdot\bOmega^{-1} &= \bX+\bY
    \label{eq:red_casida_like_xpy} \\
    (\bA+\bB)\cdot(\bA-\bB)\cdot(\bX-\bY) &= (\bX-\bY)\cdot\bOmega^2
    \label{eq:red_casida_like_xmy}
  \end{align}
\end{subequations}
So far, no assumption has been made about the positive definiteness of $\bA-\bB$.
When $\bA-\bB$ is positive definite, Eq.~\eqref{eq:red_casida_like_xmy} can be transformed into a Hermitian eigenvalue problem by multiplying it from the left by $(\bA-\bB)^{1/2}$.
We may write $\bA-\bB=(\bA-\bB)^{1/2}\cdot(\bA-\bB)^{1/2}$ and define $\bZ=(\bA-\bB)^{1/2}\cdot(\bX-\bY)$.
Equation~\eqref{eq:red_casida_like_xmy} then becomes
\begin{equation}
  (\bA-\bB)^{1/2}\cdot(\bA+\bB)\cdot(\bA-\bB)^{1/2}\cdot\bZ =\bZ\cdot\bOmega^2
  \label{eq:red_herm_casida_like_xmy}
\end{equation}
This approach is standard in real-valued implementations of RPA and avoids the use of complex arithmetic.~\cite{Marie_2024a}

For Aufbau references, $\bA-\bB$ is typically positive definite, largely because the orbital energy differences $\epsilon_a-\epsilon_i$ are positive.
For non-Aufbau references, however, this property is no longer guaranteed, since some orbital energy differences $\epsilon_a-\epsilon_i$ become negative and contribute negatively to the diagonal of $\bA$.
To halve the problem size without relying on an implementation that handles complex arithmetic, we can instead solve Eq.~\eqref{eq:red_casida_like_xmy} directly.
In this case, two additional issues arise that do not occur when $\bA-\bB$ is positive definite and the Hermitian formulation in Eq.~\eqref{eq:red_herm_casida_like_xmy} is used.

First, the eigenvectors obtained from Eq.~\eqref{eq:red_casida_like_xmy} do not automatically satisfy the appropriate RPAx normalization and must therefore be normalized after diagonalization, provided their RPAx norms are nonzero.
Second, Eq.~\eqref{eq:red_casida_like_xmy} determines only $\bOmega^2$, so the signs of the excitation energies must be recovered explicitly.
This is essential for reconstructing $\bX+\bY$, since Eq.~\eqref{eq:red_casida_like_xpy} contains the factor $\bOmega^{-1}$.
For all modes with $\Omega_\nu^2>0$, we can initially assume positive excitation energies and use Eq.~\eqref{eq:red_casida_like_xpy} to construct $\bX-\bY$ and $\bX+\bY$.
For genuine excitation modes, the RPAx eigenvectors are conventionally normalized according to
\begin{equation}
  (\bX-\bY)^\dagger\cdot(\bX+\bY)=\bI
  \label{eq:normalization_app}
\end{equation}
so that the excitation modes have positive RPAx norm, while their de-excitation partners have negative norm.~\cite{Scuseria_2008,Song_2025}
For the unnormalized eigenvectors obtained from Eq.~\eqref{eq:red_casida_like_xmy}, the matrix $(\bX-\bY)^\dagger\cdot(\bX+\bY)$ is diagonal for nondegenerate $\Omega_\nu^2$, but its diagonal elements are not yet normalized to unity.
Their signs determine the appropriate assignment of the signs of $\Omega_\nu$, since changing $\Omega_\nu\rightarrow-\Omega_\nu$ in Eq.~\eqref{eq:red_casida_like_xpy} changes the sign of the corresponding $\bX+\bY$ vector.
A negative diagonal element therefore indicates that the corresponding mode has been assigned the wrong sign of $\Omega_\nu$, in which case the associated $\bX+\bY$ vector is multiplied by $-1$.
Once the signs of the excitation energies have been assigned consistently, the eigenvectors can be normalized by applying the inverse square root of the diagonal overlap matrix $(\bX-\bY)^\dagger\cdot(\bX+\bY)$ consistently to $\bX$ and $\bY$, yielding Eq.~\eqref{eq:normalization_app}.

The reduction to Eq.~\eqref{eq:red_casida_like_xmy} and the reconstruction in Eq.~\eqref{eq:red_casida_like_xpy} are valid for modes with nonzero excitation energies.
For such modes, the reduced eigenvalue problem determines $\Omega_\nu^2$, and Eq.~\eqref{eq:red_casida_like_xpy} can be used to reconstruct $\bX_\nu+\bY_\nu$ from $\bX_\nu-\bY_\nu$.
If an exact zero mode occurs, this reconstruction is not defined because $\Omega_\nu^{-1}$ does not exist.
Moreover, Eq.~\eqref{eq:red_casida_like_xmy} then only imposes
\begin{equation}
  (\bA+\bB)\cdot(\bA-\bB)\cdot(\bX_\nu-\bY_\nu)=0
\end{equation}
whereas the original coupled RPAx equations at $\Omega_\nu=0$ require
\begin{subequations}
  \begin{align}
    (\bA+\bB)\cdot(\bX_\nu+\bY_\nu)=0 \\
    (\bA-\bB)\cdot(\bX_\nu-\bY_\nu)=0
  \end{align}
\end{subequations}
Thus, an exact zero mode cannot in general be reconstructed from Eq.~\eqref{eq:red_casida_like_xmy} alone.
Instead, $\bX_\nu+\bY_\nu$ and $\bX_\nu-\bY_\nu$ must be obtained independently from the null spaces of $\bA+\bB$ and $\bA-\bB$, respectively, and $\bX_\nu$ and $\bY_\nu$ can then be reconstructed from their sum and difference.
This restriction applies only to the zero-energy modes, and all modes with $\Omega_\nu\neq0$ can be treated using the reduced formulation above.
For an exact zero mode, the usual RPAx normalization requires additional care, since its RPAx norm, $(\bX_\nu-\bY_\nu)^\dagger\cdot(\bX_\nu+\bY_\nu)$, may vanish.
In this case, the mode cannot be normalized to the conventional unit RPAx norm and must be treated separately from the nonzero-frequency modes.

Finally, if Eq.~\eqref{eq:red_casida_like_xmy} yields a negative real eigenvalue, $\Omega_\nu^2<0$, the corresponding RPAx frequencies form a purely imaginary pair. Because the reduced matrix is non-Hermitian, it may also possess genuinely complex eigenvalues $\Omega_\nu^2$, which generate complex RPAx quartets. Both cases lie outside the real-valued reconstruction considered here.

\section{Ring-CCD in the case of complex RPAx solutions}
\label{app:cplxrCCD}

The equivalence of the ring-CCD formulation to the RPAx problem for complex-valued matrices has been presented in Ref.~\onlinecite{Kitsaras_2026}.
Using the composite particle-hole indices $ia$ and $jb$, the ring-CCD amplitudes $T_{ia,jb}=t^{ab}_{ij}$ satisfy the following algebraic Riccati equation:
\begin{equation}
	\bB +\bA \cdot \bT + \bT \cdot \bA^* + \bT \cdot \bB^* \cdot \bT  = \bO 
	\label{eq:rCCD}
\end{equation}
Provided that $\bX^*$ is invertible for the selected RPAx invariant subspace, their correspondence to the $\bX$ and $\bY$ matrices is given by
\begin{equation}
	\bT = \bY^* \cdot \left ( \bX^* \right )^{-1} 
	\label{eq:Trccd}
\end{equation}
and the RPAx correlation energy is given by
\begin{equation}
	E_\text{c} = \frac{1}{4} \Tr(\bB^* \cdot \bT)
\end{equation}

In the case of a pair of purely imaginary RPAx solutions, $\left\{\Omega_\nu,-\Omega_\nu\right\}$, our real-valued ring-CCD iterations did not converge to a real solution of Eq.~\eqref{eq:rCCD}.
To explore complex solutions, we used the complex-rotated initial guess
\begin{equation}
	\bT^{(0)}=e^{\ii\phi}\bT_\text{MP2}
\end{equation}
where $\bT_\text{MP2}$ are the (real-valued) MP2 amplitudes.
Depending on the rotation angle $\phi$, the iterations converged to either of two complex-conjugate solutions. The associated correlation energies have identical real parts, $\Re(E_\text{c})$, and imaginary parts of opposite sign, $\Im(E_\text{c})$. 
These solutions are associated with the two conjugate choices of RPA invariant subspace and yield the corresponding conjugate branches of the plasmon expression.

For a quartet of genuinely complex RPAx solutions, $\left\{\Omega_\nu,-\Omega_\nu,\Omega_\nu^*,-\Omega_\nu^*\right\}$, we were unable to converge the ring-CCD iterations from any of the complex-rotated initial guesses considered. This numerical failure does not establish that Eq.~\eqref{eq:rCCD} has no algebraic solution; it may instead reflect a singular or poorly conditioned $\bX$ block or the absence of the relevant solution from the basins of attraction sampled by our iterative procedure.

\acknowledgments{
This work used the HPC resources from CALMIP (Toulouse) under allocation 2026-18005.
}%
\section*{Data availability statement}
The data that support the findings of this study are available within the article and its supplementary material.
\section*{References}
\bibliography{RPA-instabilities.bib}
\end{document}